\documentclass[epj,nopacs]{svjour}

\input epsf
\begin{document}

\title{A DMRG study of the $q$-symmetric Heisenberg chain}
\author{M. Kaulke\thanks{\email{kaulke@physik.fu-berlin.de}} \and
I. Peschel\thanks{\email{peschel@aster.physik.fu-berlin.de}} }

\institute{Fachbereich Physik, Freie Universit\"at Berlin, Arnimallee
14, D-14195 Berlin, Germany}

\date{\today}

\abstract{
The spin one-half Heisenberg chain with $U_q[SU(2)]$ symmetry is
studied via density-matrix renormalization. Ground-state energy and
$q$-symmetric correlation functions are calculated for the
non-hermitian case $q=\exp(i\pi/(r+1))$ with integer $r$. This gives
bulk and surface exponents for (para)fermionic correlations in the
related Ising and Potts models. The case of real $q$ corresponding to
a diffusion problem is treated analytically.
\PACS{
      {PACS-key}{discribing text of that key}   \and
      {PACS-key}{discribing text of that key}
     } 
} 

\maketitle


\section{Introduction}\label{intro}
The density-matrix renormalization method (DMRG) introduced by White
\cite{white} has opened new possibilities in the study of
one-dimensional quantum systems. With this approach not only the
energies of large systems can be calculated very accurately, the
method also yields correlation functions. \cite{gehring} It was tested
e.g. on Heisenberg \cite{white} and transverse \mbox{Ising
\cite{legeza}} chains and has been applied to a variety of systems
like higher-spin chains \cite{hallberg}, \mbox{ladders \cite{white1}},
Hubbard chains \cite{qin}, impurity models \cite{wang}, phonon systems
\cite{caron} or transfer matrices \cite{nishino,wang1}. It is
therefore natural to try it in still other situations and to see how
it performs there.

The system which we have studied is the $U_q[SU(2)]$-symmetric
Heisenberg chain with Hamiltonian \cite{alcaraz3}
\begin{equation}
\label{ham}
H=\mp\sum_{i=1}^{L-1} e_i
\end{equation}
with open boundary conditions and
\begin{eqnarray}
\nonumber e_i=-\frac{1}{2} \left( \sigma_i^x \sigma_{i+1}^x +
\sigma_i^y \sigma_{i+1}^y + \frac{q+q^{-1}}{2} \left( \sigma_i^z
\sigma_{i+1}^z -1 \right) \right. \\ \left. + \frac{q-q^{-1}}{2}
\left( \sigma_i^z -\sigma_{i+1}^z \right) \right) \label{e_op}
\end{eqnarray}
where the $\sigma_i^\alpha$ are Pauli matrices. Some basic features of
the $q$-symmetry are sketched briefly in an appendix.

The motivation for considering this problem was two-fold. Firstly, the
model comprises various other physical systems and situations which
thereby can be treated in a unified way. In particular, we were
interested in calculating correlation functions and their
exponents. Secondly, the Hamiltonian is non-hermi\-tian for complex
$q$ and offers a possibility to investigate this situation.

In the antiferromagnetic case (sign $-$) and for complex
$q=\exp(i\pi/(r+1)), \,r$ integer, the
chain is related to the critical models with central charge
\mbox{$c=1-6/(r(r+1))$}, in particular the Ising
($r=3$) and the three-state Potts model ($r=5$)
\cite{alcaraz,pasquier}. For certain real values of $q$ it corresponds
to Potts models with more than three states \cite{alcaraz2}. The
ferromagnetic chain with real $q$, on the other hand, describes a
non-equilibrium problem, namely the hopping of classical hard-core
particles on discrete sites, with a bias in one direction
\cite{alcaraz1}. For $q=1$, finally, one recovers the
simple isotropic Heisenberg model.

The Hamiltonian (\ref{ham}) has real eigenvalues and can be treated by
Bethe ansatz. In this way, finite-size spectra have been obtained in
connection with conformal invariance \cite{alcaraz3}. Using the
Temperley-Lieb properties of the operators $e_i$, one can also find a
representation by hermitian IRF operators, for which some DMRG
calculations have already been done \cite{sierra}. However, as
mentioned above, we are interested in non-hermitian effects. These can
be varied in this model by changing the parameter $q$.

The general arguments for the DMRG procedure \cite{white} do not
involve the Hamiltonian and the approach is therefore applicable
here. In fact, non-hermi\-tian operators have been treated occasionally
\cite{wang1,kondev,hieida}. There is some freedom, however, which
density matrix to choose then. In our case we always used a hermitian
one, constructed from the right eigenvectors of (\ref{ham}). One then
finds that its spectrum shows a marked and systematic variation with
$r$. The eigenvalues decrease more slowly than for the simple
Heisenberg chain. Nevertheless, there is no difficulty in obtaining
the ground-state energy of $H$ to high accuracy. This is discussed in
section \ref{dmrg}.

Our main effort involved correlations. The functions which we studied
were $q$-sym\-metric generalizations of the quantity
$<\!\mbox{\boldmath$\sigma$}_{l}\mbox{\boldmath$\sigma$}_{m}\!\!>$
and correspond to fermionic and para\-fermionic correlators in the
Ising and Potts case, respectively \cite{hinr}. In a previous study
with exact diagonalizations of short systems no conclusive results
could be obtained for the Potts model \cite{arndt}. From our
calculations up to 100 sites we were able to obtain functions with
clear scaling behaviour from which bulk and surface exponents could be
extracted. The results are presented in section \ref{correlation}. In
the Ising case, where one has the exact solution as a test, they are
quite good. In the Potts case, there are still some uncertainties,
since the bulk exponents of the four functions which we considered
differ while their spinor character suggests that they are equal.

The diffusion problem, treated in section \ref{hopping}, has a very
different but also interesting nature. Here one finds that the
eigenvalues of the density matrix drop to exactly zero very rapidly,
so that the DMRG procedure terminates after a certain number of
states. This can be understood from the nature of the ground state
which is known here. From it, the density matrix can be derived and
thus one has a non-trivial example of a system where the DMRG
procedure can be carried out analytically.

A summary and some additional remarks can be found in the concluding
section \ref{conclusion}.


\section{DMRG procedure and ground-state energies}\label{dmrg}

In the DMRG approach, a selection of relevant states is made, not for
isolated, but for already interacting blocks which form the total
system. In our calculations we always used the infinite-size variant
of the algorithm \cite{white}. 

As usual the chain was divided in two parts labeled by the indices 1
and 2, respectively. Then the right ground-state vector
$|\Phi_r>=\sum\Phi_{ij}|i>_1|j>_2$, where $|i>_1$ and $|j>_2$ denote
the basis states in the two parts, and the corresponding eigenvalue
$E$ were calculated by using a combination of a vector-iteration
procedure (power method) and a Lanczos procedure with modified scalar
product. The whole calculation was done in the $S^z=0$ subspace which
reduced the numerical effort significantly.

From the ground-state vector the hermitian density matrix
\begin{equation} \label{rho_sb}
\rho=|\Phi_r><\Phi_r|
\end{equation}
of the chain was constructed where
\begin{equation}\label{eigenvec}
<\Phi_r|=|\Phi_r>^\dagger
\end{equation}
is the usual hermitian conjugate. From (\ref{rho_sb}) the reduced density matrix $\rho_1$ of part 1 of
the chain follows as
\begin{equation} \label{rho}
\rho_1 = {\displaystyle \sum \limits_{i,i',j} }
\Phi_{ij}\Phi_{i'j}^\ast |i>_{1\;1}<i'|
\end{equation}
and similarly for $\rho_2$. Since the boundary terms in (\ref{ham})
break reflection symmetry one has to work with both matrices here.

The use of hermitian density matrices is already suggested by the
considerations in \cite{white} where $\rho_1$ enters via an
optimization procedure for a vector like $|\Phi_r>$. Non-symmetric
$\rho$'s have been used in some cases \cite{nishino,wang1,hieida} but
lead to obvious problems if their eigenvalues become complex. We
tested the choice $\rho=|\Phi_r><\Phi_l|$, where $<\Phi_l|$ is the
left eigenvector of $H$, on the XY-analogue of (\ref{ham}) but it gave
unsatisfactory results, in particular large errors which did not
decrease properly with $m$.

The $m$ eigenvectors $|\lambda_n>_1, \, |\lambda_n>_2$ of $\rho_1$
and $\rho_2$ corresponding to the largest eigenvalues $\lambda_n$ are
used as new basis states for the two parts of the chain in terms of
which all necessary operators are expressed. After inserting two sites
between part 1 and part 2 the procedure is restarted.

For the calculation of expectation values 
\begin{equation} \label{exp}
<\hat O> = < \Phi_l | \hat O | \Phi_r >
\end{equation}
one also needs the left eigenvector. In the present case this is just
the transpose of $|\Phi_r>$ since $H$ is complex symmetric.

\begin{figure}[ht]
\epsfxsize=80mm
\epsffile{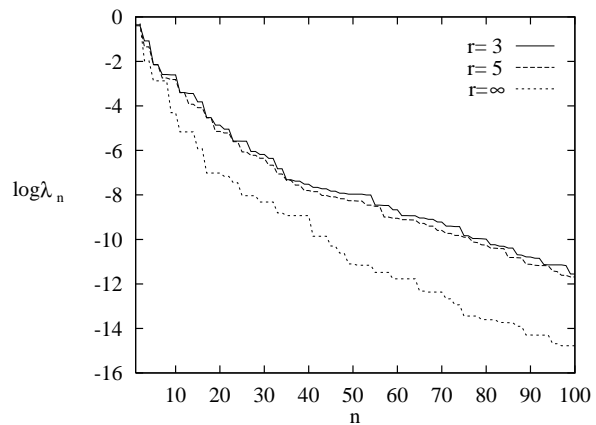}
\caption{\label{fig_spec_rho} Eigenvalue spectrum of the reduced
density matrix $\rho_1$ for different values of the parameter $r$ and
chain length $L=30$, calculated with $64$ kept states.}
\end{figure}

For the efficiency of the DMRG procedure it is important that one can
calculate accurate results with a relatively small number of states of
the Hilbert space. Therefore the eigenvalues of the reduced density
matrix should decay exponentially. However, in general it is not a
priori clear if the spectrum exhibits such a behaviour and one has to
check it numerically. Figure \ref{fig_spec_rho} shows such spectra for
different values of the parameter $r$. One can see roughly exponential
behaviour with a fast initial and a slower final decay. However, the
results for small $r$ lie considerably above those for the isotropic
Heisenberg model. The $q$-symmetric chain is therefore numerically a
less favourable case. This feature is essentially due to the imaginary
\linebreak[4] boundary terms. When added to an XX-chain they shift the
density-matrix spectrum in the same way.

\begin{table}[ht]
\begin{center}
\begin{tabular}{|c|c|c|} \hline
L & r=3 & r=5 \\ \hline\hline
 4 & -0.6604497504 & -0.7393376252  \\
10 & -0.7443482698 & -0.8012448364  \\
20 & -0.7736274980 & -0.8237779981  \\
40 & -0.7885618957 & -0.8354921842  \\
60 & -0.7935870982 & -0.8394707437  \\
80 & -0.7961088422 & -0.8414746652  \\
\hline \hline 
L &  r=7 & r=$\infty$ \\ \hline \hline
 4 &  -0.7687664240 & -0.8080127018  \\
10 &  -0.8227289432 & -0.8516070414  \\
20 &  -0.8427349544 & -0.8682473334  \\
40 &  -0.8532195604 & -0.8770736642  \\
60 &  -0.8567946053 & -0.8801004992  \\
80 &  -0.8585981414 & -0.8816309100  \\
\hline
\end{tabular}
\caption{\label{tab_ew_l} Ground state energies per spin for different
$r$, calculated with $m=128$ for $r=3,5,7$ and $m=64$ for $r=\infty$.}
\end{center}
\end{table}

We have calculated the ground state energies for different values of
the parameter $r$ and for different numbers of kept states $m$. Table
\ref{tab_ew_l} shows some results for $m=128$. (We note that the
results given here are calculated without the constant
$-(L-1)(q+q^{-1})/4$ in (\ref{ham}).) The accuracy can be checked
directly for $r=3$ by comparing with the analytic result
\cite{burkhardt}. In this case the deviations are less than $10^{-9}$
which is in accord with a truncation error of the order of $10^{-9}$
in the density matrix. One can also compare with the values obtained
by Sierra et al. \cite{sierra} in a different DMRG calculation with
$m=160$. We could reproduce their data up to 9 digits.
\begin{figure}[ht]
\epsfxsize=80mm 
\epsffile{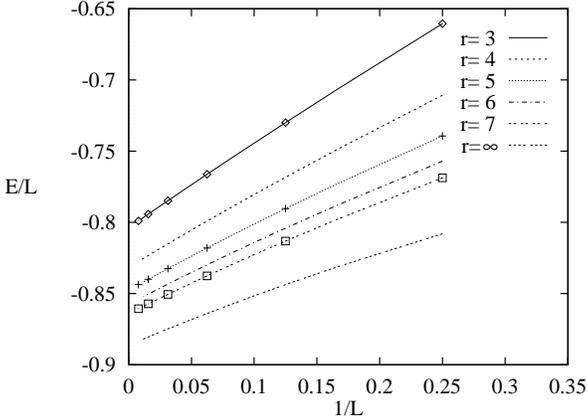}
\caption{\label{fig_ew_l} Ground state energies per site for different
values of $r$ calculated with $m=128$ (for $r=\infty$ with
$m=64$). The symbols denote data from Sierra et
al. \protect\cite{sierra}.}
\end{figure}

Plotting the ground-state energies per spin vs. $1/L$, one obtains the
well-known asymptotic behaviour
\begin{equation}\label{ew_ising}
\frac{E}{L}=\epsilon_\infty+\frac{B}{L}-\frac{C}{L^2}
\end{equation}
where $B$ is the boundary energy and $C$ measures the Casimir
effect. This is shown in figure \ref{fig_ew_l} for several values of
the parameter $r$.

\begin{figure}[ht]
\vspace{20mm}
\epsfxsize=80mm 
\epsffile{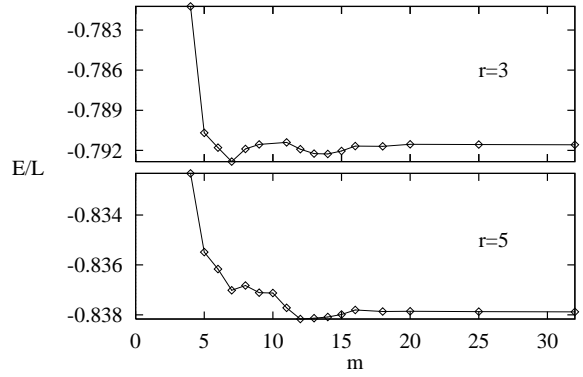}
\caption{\label{fig_ew_m} Ground-state energies per site for $r=3$ and
$r=5$ vs. the number of kept states $m$ of the density matrix for
chain length $L=50$.}
\end{figure}

One expects that for a non-hermitian operator the procedure does not
necessarily give an upper bond for $E$. This is in fact the case. For
$r=3$, the deviation from the exact result, although quite small,
changes sign for a certain $L$. Similarly, the approach of $E$ towards
its finite-size limit as a function of $m$ is not monotonous. This is
shown in \mbox{figure \ref{fig_ew_m}.} One notes, however, that the
behaviour becomes monotonous for large $r$.
%


\section{Correlation functions}\label{correlation}

For the $q$-symmetric chain with $q$ a root of unity, the usual spin
correlation functions are not the appropriate quantities. Instead, we
consider operators which are adapted to the $U_q[SU(2)]$ symmetry. In
this way one is lead to generalizations of the scalar product
$\mbox{\boldmath$\sigma$}_{l}\mbox{\boldmath$\sigma$}_{m}$ of two
spin operators given by \cite{hinr}
\begin{equation}
\label{g_op}
g_{l,m}^{\pm}=c_{l,m}^{\pm}-\frac{1}{q+q^{-1}}
\end{equation}
with
\begin{equation}
\label{c_op}
\begin{array}{lll}
c_{l,m}^{\pm} & = & b_{m-1}^{\pm} b_{m-2}^{\pm} \ldots
b_{l+1}^{\pm} e_l b_{l+1}^{\mp} \ldots b_{m-2}^{\mp}
b_{m-1}^{\mp} \\ \\
b_j^{\pm} & = & q^{\pm 1} - e_j
\end{array}
\end{equation}
where the $e_j$ are defined in equation (\ref{e_op}). Written out
explicitely, these are non-local objects containing strings of spin
operators. However, they are constructed solely in terms of the $e_j$
from which the Hamiltonian (\ref{ham}) is built. Therefore they can be
translated into any other representation of the Temperley-Lieb algebra
which the $e_j$ fulfill \cite{hinr}. In the cases $r=3$ and $5$ such
representations are given by the critical transverse Ising model and
the Potts model with $L/2$ sites and open boundary conditions. In
these cases $<g^+_{l,m}>=<g^-_{l,m}>$, thus we work in the following
only with one operator and drop the superscript.

In the Ising case one finds that the string operators cancel the
Jordan-Wigner factors in a fermion picture and the $<g_{l,m}>$
become simple fermionic two-point functions \cite{hinr}. For example
\begin{eqnarray}
<g_{2j-1,2k}> & = & -\frac{1}{\sqrt{2}}(-1)^{j+k}\nonumber \\ & &
\times \left<\left(d^\dagger_j+d_j\right)
\left(d_k^\dagger-d_k\right)\right>
\label{g_op_ising_1} \\ 
<g_{2j,2k-1}> & = & -\frac{1}{\sqrt{2}}(-1)^{j+k} \nonumber \\ & &
\times \left<\left(d^\dagger_j-d_j\right)
\left(d_k^\dagger+d_k\right)\right>
\label{g_op_ising_2} 
\end{eqnarray}
where the $d_j,\,d_j^\dagger$ are fermion operators. These functions
can be calculated exactly and are given as sums for finite $L$. The
two quantities $<g_{2j,2k}>$ and $<g_{2j-1,2k-1}>$ vanish
identically.  The other two correspond to different behaviour at the
boundaries where (\ref{g_op_ising_1}) remains finite and
(\ref{g_op_ising_2}) vanishes for fixed $k$ and finite chain
lengths. Due to the open ends, one is dealing with surface critical
behaviour here. After performing the continuum limit one can determine
the critical exponents, and one finds the bulk exponent $x=1/2$ for
both correlation functions and the surface exponents $x_s=1/2$ and
$x_s=3/2$ for (\ref{g_op_ising_1}) and (\ref{g_op_ising_2}),
respectively \cite{hinr}.

For the Ising case one can also construct the conventional spin
correlation function $<\sigma^x_l\sigma^x_m>$ from the quantities
$e_j$ appearing in $H$, using the property
\mbox{$\left(\sigma^x\right)^2=1$.} However, this is not possible for
the Potts model. Also order parameter profiles, as determined by a
direct DMRG calculation in \cite{igloi}, cannot be obtained in the
present formulation. However, the $g$'s are also interesting
quantities since they become parafermion operators in this case
\cite{hinr,mittag}. Their exponents will be studied in the
following.

We considered the correlation function $<g_{i,L/2}>$
($i=1,\ldots,L/2-1$) with one point fixed in the middle of the chain
while varying the position of the other one. In this way profiles were
computed for chains with up to 100 sites. The correlations were
significantly less accurate than the energy, so that in general
$m=200$ states were kept. This was a practical limit with respect to
storage and time. A complete run for fixed $r$ took several weeks CPU
time on middle-performance DEC-Alpha workstations. The amount of
memory was about 250-300 MB RAM and up to 500 MB harddisk space.

\begin{figure}[ht]
\epsfxsize=80mm 
\epsffile{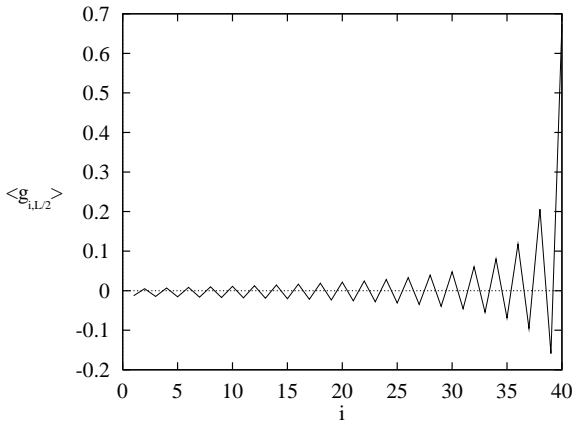}
\caption{\label{fig_gp_i}Correlation function $<g_{i,L/2}>$ for
$r=5$ and chain length $L=82$, calculated with $200$ kept states.}
\end{figure}

An example of the resulting correlation function is shown in figure
\ref{fig_gp_i}. The oscillations are due to the antiferromagnetic
character of the chain. Except for the case $r=3$, where only the two
functions (\ref{g_op_ising_1}) and (\ref{g_op_ising_2}) are non-zero,
one can distinguish four different functions given by the maxima and
minima for even and odd $L/2$.

\begin{figure}[ht]
\epsfxsize=80mm 
\epsffile{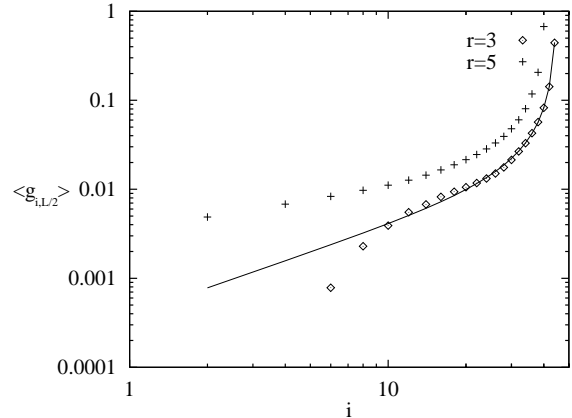}
\caption{\label{fig_gp_i_log}Log-log plot of the correlation function
$<g_{i,L/2}>$ for even $i$ and chain length $L=82 \; (r=5)$ and
$L=90 \;(r=3)$, calculated with 200 kept states. The curve
denotes analytic results for $r=3$.}
\end{figure}

As a check on the accuracy, one can compare with exact results for
$L=24$ \cite{heinzel}. One then finds deviations less than $10^{-4}$
in the Ising case and less than $10^{-5}$ in the Potts case while the
truncation error is only $10^{-14}-10^{-15}$. The correlations are
therefore much less precise than the ground-state energy. One also
finds that the values can be above or below the exact result. The
errors increase, as usual in the DMRG, as one moves towards the
boundary. This effect can also be seen in figure \ref{fig_gp_i_log},
where two correlation functions are shown for large systems. Up to
$R\simeq 30$, where $R=L/2-i$ is the distance from the center, the
numerical values for $r=3$ reproduce the analytical ones quite
well. Near the boundary, where the function is very small, however,
there are considerable deviations and the expected power-law behaviour
$g\sim R_s^{\,x_s-x}$, with $R_s=i$ denoting the distance from the
surface, breaks down. This is different for $r=5$, where the curve
remains linear also in this region. One therefore expects that the
results for larger $r$ are more accurate.

\begin{figure}[ht]
\epsfxsize=80mm 
\epsffile{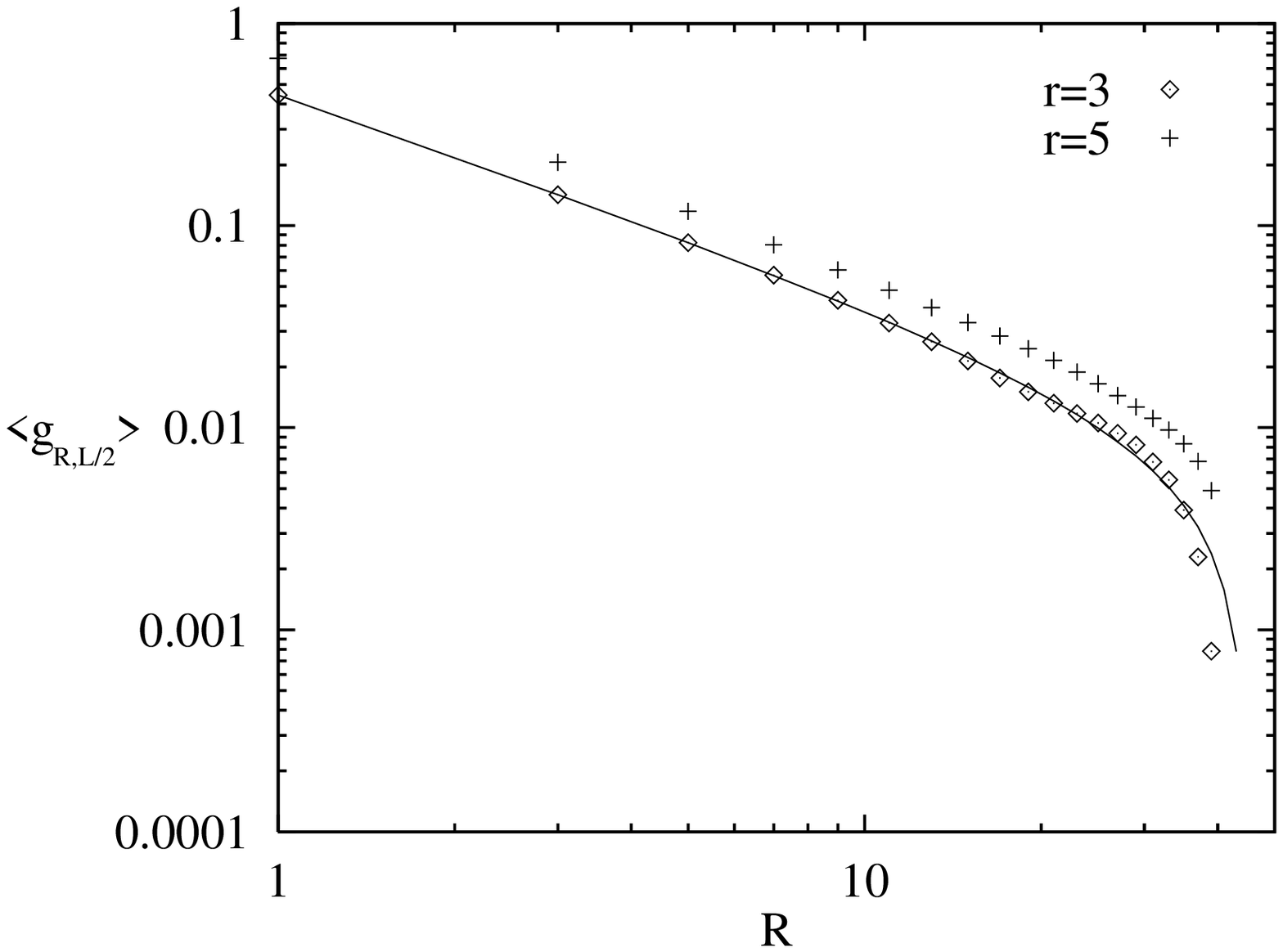}
\caption{\label{fig_gp_rbulk_log}Log-log plot of the correlation
function \mbox{$<g_{R,L/2}>$,} $R=L/2-i$, for odd $R$ and chain length
$L=82 \; (r=5)$ and $L=90 \; (r=3)$, calculated with 200 kept
states. The curve denotes analytic results for $r=3$.}
\end{figure}

In the middle of the chain, the situation is more favour\-able since
the quantities have undergone fewer iterations here. For small $R \;
(R\ll L/2)$ one finds the expected bulk behaviour $g\sim R^{-2x}$, as
shown in the log-log plot of figure \ref{fig_gp_rbulk_log}. However,
if one determines the exponent $x$ from this data for $r=3$, one still
has a relative error of about 5\%.

\begin{figure}[ht]
\epsfxsize=80mm 
\epsffile{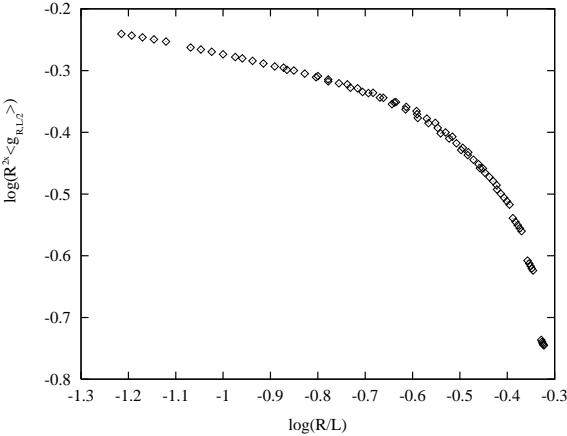}
\caption{\label{fig_fss_rbulk}Finite-size scaling behaviour of the
correlation function \protect\mbox{$<g_{R,L/2}>,$} $R=L/2-i$, for
$r=5$ and five chain lengths from $L=66$ to $L=82$, calculated with
$200$ kept states.}
\end{figure}

In order to improve the accuracy, we used the scaling form for $g$
\begin{equation}\label{scaling-bulk}
g(R,L)=\frac{1}{R^{\,2x}}F\left(\frac{R}{L}\right)
\end{equation}
where $F(R/L)\rightarrow const$ for $R/L\rightarrow 0$. Collecting
results for different $L$ we constructed a scaling plot $R^{\, 2x}g$
vs. $R/L$ and varied $x$ until the data fell onto a curve as well as
possible. The result of such a procedure for $r=5$ is shown in figure
\ref{fig_fss_rbulk}. It proves that scaling is indeed fulfilled, in
contrast to \cite{arndt} where it could not be seen for chains up to
$L=28$. The exponent $x$ for $r=3$ then is only about 2\% off the
exact result $x=1/2$.

A further improvement could be achieved by extrapolating the bulk
values of $g$ to the infinite-chain limit. Then a linear fit of the
data in the corresponding plot gave better Ising exponents. In this
way, the results in table \ref{tab_exp} were determined. The
confidence interval of the fit parameters would give rise to an
absolute error in the exponents less than $10^{-3}$, but the error in
the data points is not known. Thus we cannot give a reasonable error
bar but we can use the results for $r=3$ as a clue.

\begin{table}[ht]
\begin{tabular}{|c|c|c|c|c|} \hline
\rule[-3mm]{0mm}{8mm} r &$\!\!<g_{2j,2k}>\!\!$& 
$\!\!<g_{2j,2k-1}>\!\!$& 
$\!\!<g_{2j-1,2k}>\!\!$& 
$\!\!<g_{2j-1,2k-1}>\!\!\!$\\ 
\hline\hline
3 & - & 0.499 & 0.502 & - \\ \hline 
4 & 0.386 & 0.521 & 0.513 & 0.406 \\ \hline
5 & 0.397 & 0.520 & 0.508 & 0.409 \\ \hline
6 & 0.392 & 0.516 & 0.502 & 0.417 \\ \hline
7 & 0.412 & 0.499 & 0.484 & 0.422 \\ \hline
\end{tabular}
\caption{\label{tab_exp} Critical exponents $x$ of the four
correlation functions, calculated with $m=200$. Exact results are
known for $r=3$ and $r=\infty$ where $x=1/2$.}
\end{table}

The final Ising exponents differ only by $10^{-3}$ from the exact
value and thus are very good. The situation is more complicated,
though, for other cases. The Kac table \cite{christe} for $r=5$
contains 10 different conformal dimensions and allows for a large
number of combinations $x=\Delta + \bar{\Delta}$. However, if one
fixes the spin $s$ of the operators to simple values, only $x=11/20 \;
(s=1/2)$, $x=7/15 \; (s=1/3)$ and $x=2/3 \; (s=2/3)$ remain. These
values were also obtained directly by relating the three-state Potts
model to a Gaussian model \cite{nienhuis}. From the explicit form of
the parafermion operators (products of local and string operators)
\cite{hinr}, it can be seen that they have spinor properties as
discussed in \cite{nienhuis} with $s=1/3$. Thus one would expect a
common exponent $x=0.467$ for all four $g$'s, whereas the numerical
values lie about 10\% higher and lower. In principle, they would allow
also other $x$-values connected with strange spin. A similar situation
is found for $r=4$ (tricritical Ising model) where $x=7/16 \;
(0.438)$, $x=19/40 \;(0.475)$ and $x=43/80 \; (0.537)$ are close to
the measured values.

The reason for the discrepancies is not clear. As noted above, one
would expect better accuracy for the Potts model. If the $x$-values
actually differ, it could perhaps be due to a cancellation of leading
singularities in the $g$'s which, when written in the Potts
representation, are sums of two contributions. The other possibility
is that in spite of the satisfactory scaling behaviour the numerical
results are not yet good enough.

We also determined the surface exponents $x_s$. This was done with the
help of another scaling plot. For $i=R_s$, the scaling form of $g$ can
be written as
\begin{equation}\label{scaling-surf}
R_s^{\,2x}\,g(R_s,L)=
\left(\frac{R_s}{L}\right)^{x_s+x}G\left(\frac{R_s}{L}\right)
\end{equation}
where $G\rightarrow const$ for $R_s/L\rightarrow 0$. Using the bulk
exponents as determined above, one can now find the surface exponent
by tuning $x_s$ until the scaling behaviour (\ref{scaling-surf}) is
fulfilled. In this procedure the first few sites near the boundary,
for which the errors are large, were left out.

For the Ising case we found $x_s=1.58$ for the function
(\ref{g_op_ising_1}) and $x_s=0.56$ for the function
(\ref{g_op_ising_2}). One notes that these results differ more from
the exact values than the bulk results.  For the Potts case the
following exponents were found: $x_s=0.81$ for $<g_{2j,2k}>$,
$x_s=1.05$ for \mbox{$<g_{2j,2k-1}>$,} $x_s=0.63$ for $<g_{2j-1,2k}>$ and
$x_s=0.55$ for $<g_{2j-1,2k-1}>$.

We have also determined the exponents for higher $r$. The values tend
to their Heisenberg limits $x=1/2$ and $x_s=1$ for $r\rightarrow
\infty$. However, while the exponents of the function $<g_{2j,2k-1}>$
approach this limit from above, the exponents of the other functions
always lie below it. Furthermore, the limiting values are not easy
to obtain due to the logarithmic corrections which occur for $q=1$
\cite{hallberg1}.


\section{$q$-symmetric driven diffusion}\label{hopping}

It is well-known that the isotropic ferromagnetic Heisenberg model
also describes (in all dimensions) the hopping of classical particles
on a lattice with the exclusion of double occupancy \cite{alex}. This
follows by expressing the master equation for the probability vector
$|P>$ describing the system
\begin{equation}\label{master}
\frac{\partial}{\partial t}|P> = -H|P>
\end{equation}
in a spin one-half language where a spin up (down) corresponds to
an occupied (empty) site. The stationary state is therefore the
ferromagnetic ground state in the corresponding sector of fixed $S^z$.

A similar result holds if there is a bias in the hopping. In one
dimension and after a canonical transformation the time-evolution
operator then takes exactly the form (\ref{ham}) with $q$ given by
$q=(\alpha_-/\alpha_+)^{1/2}$ where $\alpha_+ \, (\alpha_-)$ is the
hopping rate to the right (left) \cite{alcaraz1}. The stationary
properties of this diffusion problem have already been studied
\cite{sandow}. Due to the bias, the particles accumulate at one end,
thus producing a non-trivial density profile. In the magnetic
language, one is dealing with a uniaxial ferromagnet to which opposite
boundary fields are applied at the two ends. These are real and $H$ is
hermitian in this case.

\begin{figure}[ht]
\epsfxsize=80mm 
\epsffile{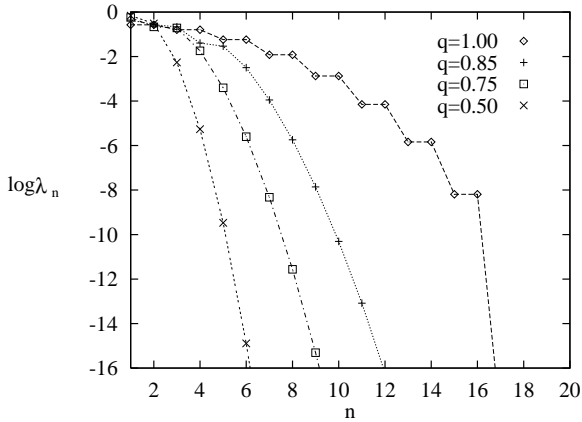}
\caption{\label{fig_spec_rho_hop}Eigenvalue spectrum of the reduced
density matrix $\rho_1$ of the $q$-symmetric hopping model at
half-filling for different values of the parameter $q$, calculated
with $16$ kept states and chain length $L=30$.}
\end{figure}

The numerical treatment leads to a density-matrix \linebreak[4]
spectrum as shown in figure \ref{fig_spec_rho_hop}. The eigenvalues
decrease faster than
exponentially and become effectively zero beyond a certain point
depending on the size of the system. The decrease also depends on the
asymmetry parameter $q$.

To understand this, it is useful to consider the case $q=1$
first. Then the ground state of the chain is a spin state $|S,M>$
where $S=L/2$ and $M=S^z$ can be chosen. Taking $M=0$ and dividing the
chain into two halves, it can be written as
\begin{equation}\label{gs_hop}
|\Phi>=|S,0>=\sum_{m=-s}^{+s}c_m |s,m>_1 |s,-m>_2
\end{equation}
where $|s,m>_{1,2}$ are the corresponding spin states of the two
subsystems with $s=S/2=L/4$ and the $c_m$ are the appropriate
Clebsch-Gordan coefficients. Thus $|\Phi>$ is the superposition of
$(S+1)$ terms and the density matrix $\rho_1$ of part 1 of the chain
has only that many non-zero eigenvalues. It reads explicitely
\begin{equation}\label{rho_hop}
\rho_1=\sum_m |s,m>_1c^2_{m\;\,1}\!\!<s,m|
\end{equation}
so that these eigenvalues are given by $c^2_m$.

For $q\ne 1$ the situation does not change, because there are
analogous decomposition formulae for the $q$-deformed angular-momentum
states \cite{kirillov}. In this case one finds
\begin{equation}\label{clebsch}
c_m=\frac{1}{\sqrt{[4s]!}}\frac{[2s]!^{\,2}}{[s-m]!\,[s+m]!}\,q^{2sm}
\end{equation}
with the notation
\begin{equation}\label{q-def}
[n]=\frac{q^n-q^{-n}}{q-q^{-1}}
\end{equation}
and
\begin{equation}\label{q-fac}
[n]!=[1]\cdot[2]\cdot\ldots\cdot[n].
\end{equation}
For $q\rightarrow 1$, $[n]$ reduces to $n$ and (\ref{clebsch}) becomes
the normal Clebsch-Gordan coefficient. One can check
that the $c^2_m$ given by (\ref{clebsch}) are precisely the
density-matrix eigenvalues obtained in the numerical procedure.

Due to the $q$-factor in (\ref{clebsch}) there is no symmetry between
$m$ and $-m$ for $q\ne 1$ and the degeneracy of the eigenvalues is
lifted as seen in figure \ref{fig_spec_rho_hop}. Specifically for
large $q$ one has
\begin{equation}\label{clebsch_lim}
c_m\simeq q^{-\left(s-m\right)^2}
\end{equation}
so that the coefficient $c_s\simeq 1$ dominates and $|\Phi>$ becomes
approximately a product state
\begin{equation}\label{gs_hop_prod}
|\Phi>\simeq |s,s>_1|s,-s>_2.
\end{equation}
Physically this means that one half of the chain is almost filled with
particles and the other one is almost empty, as expected for large
bias. Expressions for the density profile were given in \cite{sandow}
and are easily reproduced by the DMRG calculations. Similarly, the
treatment can be extended to $M\ne 0$, i.e. to systems which are not
half-filled. Then the number of terms in $|\Phi>$ is even smaller,
namely $(S+1-|M|)$, and becomes one for a completely full or empty
system. The DMRG procedure gives exact results once $m$ exceeds this
value which increases at most linearly with the system size.


\section{Conclusion}\label{conclusion}

We have studied the $q$-symmetric Heisenberg chain for various cases
of physical interest. For the case of complex $q$ we showed that the
DMRG procedure works well even though the density-matrix spectra are
less favourable due to the non-hermitian boundary terms. We found the
ground-state energies to high accuracy and calculated also generalized
correlation functions, for which we determined the critical
exponents. 

The study of various different models via the spin one-half operator
$H$ was also motivated numerically, since the dimension of the
matrices in the DMRG procedure is lower then. The results show,
however, that to some extent this advantage is compensated by the
sensivity of the correlations in this formulation. One should also
note that the equivalent Potts chain has only half the length of the
$q$-symmetric chain. Thus it seems that one cannot really circumvent
certain features of the corresponding problems. To improve the results
further, one would have to increase the numerical effort, by using
larger values of $m$ or the sweeping procedures in the finite-size
algorithm \cite{white}.

The ferromagnetic chain for real $q$ was seen to have very different
features and is interesting in other respects. Firstly, as an example
where the density matrix can be found analytically and the DMRG
procedure automatically gives the exact result. There are only a few
other cases of this kind, for example two coupled oscillators
\cite{han} or finite-dimensional matrix-product states
\cite{klumper}. Secondly, it describes not only a magnetic problem,
but also a non-equilibrium system. The use of the DMRG in this field,
where the time evolution operators are in general non-hermitian, has
only started \cite{hieida} and further interesting applications can be
expected.

\section*{Appendix}
\setcounter{equation}{0}
\renewcommand{\theequation}{\mbox{A.\arabic{equation}}}

The $q$- or quantum group symmetry (see e.g. \cite{pasquier}) is a
generalization of rotational symmetry and characterized by a modified
commutator
\begin{equation}
\left[ S^+,S^- \right]=\left[ 2S^z \right]
\end{equation}
between the generators. Here the bracket is defined in (\ref{q-def})
so that a $q$-dependent function of $S^z$ appears on the right. For
$q=1$, the usual angular momentum algebra is recovered. For a chain of
spins, $S^z=\sum \sigma^z_n/2$ has the usual form, but $S^\pm$ are
given by
\begin{equation}
S^\pm=\frac{1}{2}\sum_{n=1}^{L}
q^{\,\sum\limits_{l=1}^{n-1}\sigma^z_l/2}
\sigma_n^\pm \,
q^{-\!\!\!\sum\limits_{l=n+1}^{L}\sigma^z_l/2}.
\end{equation}

The property $\left[ H,S^\pm \right]=0$ of the Hamiltonian (\ref{ham})
can be used, as in the case $q=1$, to obtain all ferromagnetic ground
states from the one with all spins up via repeated application of
$S^-$ \cite{sandow}. In this way one can also derive the
Clebsch-Gordan coefficients (\ref{clebsch}).

\section*{Acknowledgments}

We would like to thank H. Hinrichsen, V. Rittenberg, F. Igl\'oi,
M. Henkel and H. Niggemann for useful discussions and P. Arndt and
T. Heinzel for correspondence. We furthermore thank X. Wang for his
advice in the DMRG method, A. Honecker for help with the numerics, the
Universit\'e Henri Poincar\'e, Nancy, for hospitality and the
\linebreak[4] Max-Planck-Institut f\"ur Physik komplexer Systeme in
\linebreak[4] Dresden for substantial computer time.



\begin{thebibliography}{99}
\bibitem{white}
S.R. White, Phys. Rev. Lett. {\bf 69}, (1992) 2863; S.R. White,
Phys. Rev. B {\bf 48}, (1993) 10345.

\bibitem{gehring} For a short review see G.A. Gehring, R.J. Bursill and
T. Xiang, Acta Physica Polonica A {\bf 91}, (1997) 105.

\bibitem{legeza}
\"O. Legeza and G. F\'ath, Phys. Rev. B {\bf 53}, (1996) 14349.

\bibitem{hallberg}
U. Schollw\"ock and T. Jolicoeur, Europhys. Lett. {\bf 30}, (1995) 493;
K. Hallberg, X.Q.G. Wang, P. Horsch and A. Moreo,
Phys. Rev. Lett. {\bf 76}, (1996) 4955.

\bibitem{white1}
S.R. White, R.M. Noack and D.J. Scalapino, Phys. Rev. Lett. {\bf 73},
(1994) 886; K. Hida J. Phys. Soc. Jap. {\bf 64}, (1995) 4896;
S.R. White and D.J. Scalapino Phys. Rev. B {\bf 55}, (1997) R14701; B
{\bf 57}, (1998) 3031.

\bibitem{qin}
S. Qin, S. Liang, Z. Su and L. Yu, Phys. Rev. B {\bf 52}, (1995) R5475.

\bibitem{wang}
X. Wang and S. Mallwitz, Phys. Rev. B {\bf 53}, (1995) R492; X. Wang,
cond-mat/9705302, (1997).

\bibitem{caron}
L.G. Caron and S. Moukouri, Phys. Rev. Lett. {\bf 76}, (1996) 4050;
Phys. Rev. B {\bf 56}, (1997) R8471.

\bibitem{nishino}
T. Nishino, J. Phys. Soc. Jap. {\bf 64}, (1995) 3598; T. Nishino and K. Okunishi, J. Phys. Soc. Jap. {\bf
66}, (1997) 3040.

\bibitem{wang1}
R.J. Bursill, T. Xiang and G.A. Gehring, J. Phys.: Cond. Mat. {\bf
40}, (1996) L583; X. Wang and T. Xiang, Phys. Rev. B {\bf 56}, (1997)
5061; N. Shibata, J. Phys. Soc. Jap. {\bf 66}, (1997) 2221.

\bibitem{alcaraz3}
F.C. Alcaraz, M.N. Barber and M.T. Batchelor, Phys. Rev. Lett. {\bf 58},
(1987) 771; F.C. Alcaraz, M.N. Barber, M.T. Batchelor, R.J. Baxter and
G.R.W. Quispel, J. Phys. A: Math. Gen. {\bf 20}, (1987) 6397.

\bibitem{alcaraz}
F.C. Alcaraz, M. Baake, U. Grimm and V. Rittenberg,
J. Phys. A {\bf 22}, (1989) L5.

\bibitem{pasquier}
V. Pasquier and H. Saleur, Nucl. Phys. B {\bf 330}, (1990) 523.

\bibitem{alcaraz2}
F.C. Alcaraz, M.N. Barber and M.T. Batchelor, Ann. Phys. {\bf 182},
(1988) 280.

\bibitem{alcaraz1}
F.C. Alcaraz, M. Droz, M. Henkel and V. Rittenberg, Ann. Phys. {\bf
230}, (1994) 250.

\bibitem{sierra}
G. Sierra and T. Nishino, Nucl. Phys. B {\bf 495}, (1997) 505.

\bibitem{kondev}
J. Kondev and J.B. Marston, Nucl. Phys. B {\bf 497}, (1997) 639.

\bibitem{hieida}
Y. Hieida, J. Phys. Soc. Jap. {\bf 67}, (1998) 369.

\bibitem{hinr}
H. Hinrichsen, P.P. Martin, V. Rittenberg and M. Scheu\-nert,
Nucl. Phys. B {\bf 415}, (1994) 533.

\bibitem{arndt}
P.F. Arndt and T. Heinzel, J. Phys. A: Math. Gen. {\bf 28}, (1995) 3567.

\bibitem{burkhardt}
T.W. Burkhardt and I. Guim, J. Phys. A: Math Gen. {\bf 18}, (1985) L33.

\bibitem{igloi}
E. Carlon and F. Igl\'oi, cond-mat/9710144, (1997).

\bibitem{mittag}
L. Mittag and M.J. Stephen, J. Math. Phys. {\bf 12}, (1971) 441.

\bibitem{heinzel}
T. Heinzel, \textit{Diplomarbeit} (Bonn 1995).

\bibitem{christe}
P. Christe and M. Henkel, \textit{Introduction to Conformal Invariance and
Its Applications to Critical Phenomena} (Springer, Berlin 1993).

\bibitem{nienhuis}
B. Nienhuis and H.J.F. Knops, Phys. Rev. B {\bf 32}, (1992) 1872.

\bibitem{hallberg1}
K.A. Hallberg, P. Horsch and G. Mart\'\i nez, Phys. Rev. B {\bf 52},
(1995) R719.

\bibitem{alex}
S. Alexander and T. Holstein, Phys. Rev. B {\bf 18}, (1978)
301; W. Dieterich, P. Fulde and I. Peschel, Adv. in Physics {\bf 29},
(1980) 527.

\bibitem{sandow}
S. Sandow and G. Sch\"utz, Europhys. Lett. {\bf 26}, (1994) 7.

\bibitem{kirillov}
A. Kirillov and N.Y. Reshetikhin in \textit{Infinite Dimensional Lie
Algebras and Groups} ed. V.G. Kac (World Scientific, Singapore 1989).

\bibitem{han}
D. Han, Y.S. Kim and M.E. Noz, cond-mat/9705029, (1997).

\bibitem{klumper}
see, e.g. A. Kl\"umper, A. Schadschneider and J. Zittartz,
Europhys. Lett. {\bf 24}, (1993) 293; F. E{\ss}ler and V. Rittenberg,
J. Phys. A: Math. Gen. {\bf 29}, (1996) 3375.
\end{thebibliography}
\end{document}